\documentclass[prd,superscriptaddress,twocolumn,preprintnumbers,%
  amsmath,amssymb,longbibliography]{revtex4-1}
  
\usepackage[breaklinks,colorlinks=true]{hyperref}
\usepackage[T1]{fontenc}
\usepackage[utf8]{inputenc}
\usepackage{mathtools}
\usepackage{amsmath,amsthm,amssymb}
\usepackage{amsmath,bm}
\usepackage{bbold} 
\usepackage{dsfont}
\usepackage{lipsum}
\usepackage{calrsfs}
\DeclareMathAlphabet{\pazocal}{OMS}{zplm}{m}{n}

\usepackage{color}
\allowdisplaybreaks

\usepackage[dvipsnames]{xcolor}
\newcommand\myshade{85}
\colorlet{mylinkcolor}{BrickRed}
\colorlet{mycitecolor}{NavyBlue}
\colorlet{myurlcolor}{Aquamarine}
\hypersetup{
  linkcolor  = mylinkcolor!\myshade!black,
  citecolor  = mycitecolor!\myshade!black,
  urlcolor   = myurlcolor!\myshade!black,
  colorlinks = true,
}

\newcommand{\dd}{\mathrm{d}}

\begin{document}
\title{Geometric Structures Induced by Deformations of the Legendre Transform}
\author{Pablo A. Morales}
\affiliation{Research Division, Araya Inc., Tokyo 107-6019, Japan}
\email{pablo$_$morales@araya.org}
\author{Jan Korbel}
\affiliation{Section for Science of Complex Systems, Center for Medical Data Science, Medical University of Vienna, Spitalgasse, 23, 1090 Vienna, Austria}
\affiliation{Complexity Science Hub Vienna, Josefst\"{a}dter Strasse 39, 1080 Vienna, Austria}
\author{Fernando E. Rosas}
\affiliation{Department of Informatics, University of Sussex, Brighton BN1 9RH, UK}
\affiliation{Centre for Psychedelic Research, Department of Brain Science, Imperial College London, London SW7 2DD, UK}
\affiliation{Centre for Complexity Science, Imperial College London, London SW7 2AZ, UK}
\affiliation{Centre for Eudaimonia and Human Flourishing, University of Oxford, Oxford OX3 9BX, UK}

\newtheorem{definition}{Definition}
\newtheorem{theorem}{Theorem}
\newtheorem{lemma}{Lemma}
\newtheorem{proposition}{Proposition}
\newtheorem{corollary}{Corollary}
\newtheorem{example}{Example}
\newtheorem{remark}{Remark}

\begin{abstract}
The recent link discovered between generalized Legendre transforms and non-dually flat statistical manifolds suggests a fundamental reason behind the ubiquity of R\'{e}nyi's divergence and entropy in a wide range of physical phenomena. However, these early findings still provide little intuition on the nature of this relationship and its implications for physical systems. Here we shed new light on the Legendre transform by revealing the consequences of its deformation via symplectic geometry and complexification. These findings reveal a novel common framework that leads to a principled and unified understanding of physical systems that are not well-described by classic information-theoretic quantities.
\end{abstract}

\maketitle

\section{Introduction}

The Legendre transform~\cite{rockafellar1997convex} plays a key---albeit perhaps not always transparent---role in many areas of mathematical physics. 
Specifically, it allows for the identification of dual coordinates and potentials that yield theories in terms of more convenient variables, being instrumental in diverse areas in physics ranging from 
relativistic field theory to condensed matter physics. 
Applications of the transform have their roots in classical physics---in analytical mechanics serving as a link between its Lagrangian and Hamiltonian formulations, and in thermodynamics bridging intensive and extensive variables. These notions have led to more general frameworks which, in turn, gave rise to the development of symplectic topology~\cite{mcduff2017introduction}. 

Far from being a relic, the Legendre transform still plays an important role in contemporary physics. 
It plays an important role in classical field theory, where the index of pairs of components becomes continuous. It is also used in quantum field theory, where it relates the generator of connected Green functions to the quantum effective action, i.e., the generator of one-particle irreducible Green functions. 
Furthermore, the relevance of the Legendre transform has lead to generalizations in the context of perturbative quantum field theories~\cite{jackson2017robust,krupkova2001legendre}. Overall, the transform continues to be at the core of important developments in current research.

The Legendre transform also plays a fundamental role in information geometry, where it mediates the relationship between primal and dual coordinates within the non-Riemannian geometry induced by dually flat statistical manifolds~\cite{amari2016information}. This duality gives rise to relationships of orthogonality in these geometries, corresponding to alternative representations of physical systems based on control parameters or expectation values~\cite{amari2001information}. Interestingly, the generalized Legendre transform naturally arises in curved (i.e., non-Euclidean) statistical manifolds~\cite{ohara2009geometric,scarfone2018information}, which establishes a rigorous and highly non-trivial link with R\'{e}nyi's divergence and entropy~\cite{wong2018logarithmic,PhysRevResearch.3.033216,Wong9733874}. 
These recent findings suggest the existence of a fundamental reason that could explain why R\'{e}nyi entropy and divergence naturally appear in a range of physical phenomena of interest. 
In effect, recent applications of R\'{e}nyi measures to physics includes quantum systems~\cite{stephan2014geometric,ShannonRenyiQuantumSpin}, strongly coupled or entangled systems~\cite{Dong:2016fnf,Barrella:2013wja,jizba2019maximum}, phase transitions~\cite{GeometricMutInf,DetectingPhaseTwithRenyi,PhysRevLett.107.020402}, and multifractal \mbox{thermodynamics \cite{jizba2004,jizba04b},} among others. 
However, these early findings on the link between the generalized Legendre transforms and curved geometries still provide little intuition on the nature of this relationship and its meaning and implications for physical systems in general. 

The goal of this article is to shed new light on the generalized Legendre transform by investigating its geometric implications. 
For this purpose, we characterize deformations in the Legendre transform and relate them with generalizations of the Bregman divergence, which are naturally associated with curved statistical manifolds. 
By leveraging these tools, our contribution focuses on two domains: geometrical aspects related to phase-space flow and manifold complexification. 
Our results show how the symplectic structure induced by the deformed Legendre transforms leads to a modification of what is understood as a `canonical pair,' which in turn illuminates the nature of the corresponding maximum entropy distributions. Furthermore, our results bring new insights to the relationship between the Kullback--Leibler divergence (related to the Shannon entropy), $\alpha$-divergence (related to Tsallis' entropy), and the R\'{e}nyi divergence via manifold complexification and  {K\"{a}hler} manifolds. 
The complex geometry yields new conditions on the possible values of the manifold curvature, which are closely related to holomorphic polarization. 
Additionally, we report on the thermodynamic aspects related to the deformed Legendre transform in Ref.~\cite{thermodynamicsKolmogorov}. Taken together, these results lead to a larger, unified picture that extends standard geometric and thermodynamic relationships associated with classic information-theoretic quantities such as Shannon's entropy.

The rest of the paper is structured as follows. First, Section~\ref{sec:prelim} provides a brief overview on the standard interpretation of the Legendre transform in mathematical physics. Then, Section~\ref{sec:legendre_in_geo} explores how the transform naturally arises in information geometry and introduces the intimate relationship that exists between a generalized Legendre transform and the curvature of statistical manifolds. Building on these foundations, Section~\ref{sec:sym_and_kahler} investigates the consequences of generalized Legendre transforms on the symplectic structure and flows and on the complexification of statistical manifolds. Finally, Section~\ref{sec:conclusion} summarizes our mains conclusions.

\section{Preliminaries}
\label{sec:prelim}

The Legendre transform is, at its core, an exploration of the properties of convex functions. Despite its importance, the transform is unfortunately typically introduced as an obscure algebraic `trick', with no explanation of why it plays such an important role in many different areas of physics. 
For completeness, this section presents a basic standard interpretation of the Legendre transform in mathematical physics, which is then complemented by a deeper view based on information geometry in Section~\ref{sec:legendre_in_geo}.

The most straightforward interpretation of the Legendre transform comes from the geometry of graphs of functions~\cite{zia2009making}. In this view, the Legendre transform of a convex function $F$ is another function $G$ that keeps track of the (negative) height at which the tangent to $F$ touches the y-axis, which is usually reparametrized in terms of the slope of $F$. This view is easy to grasp, but unfortunately makes the construction seem arbitrary while failing to explain why this procedure is so fundamental.

\begin{figure*}[ht]
    \centering
    \includegraphics[height=6cm]{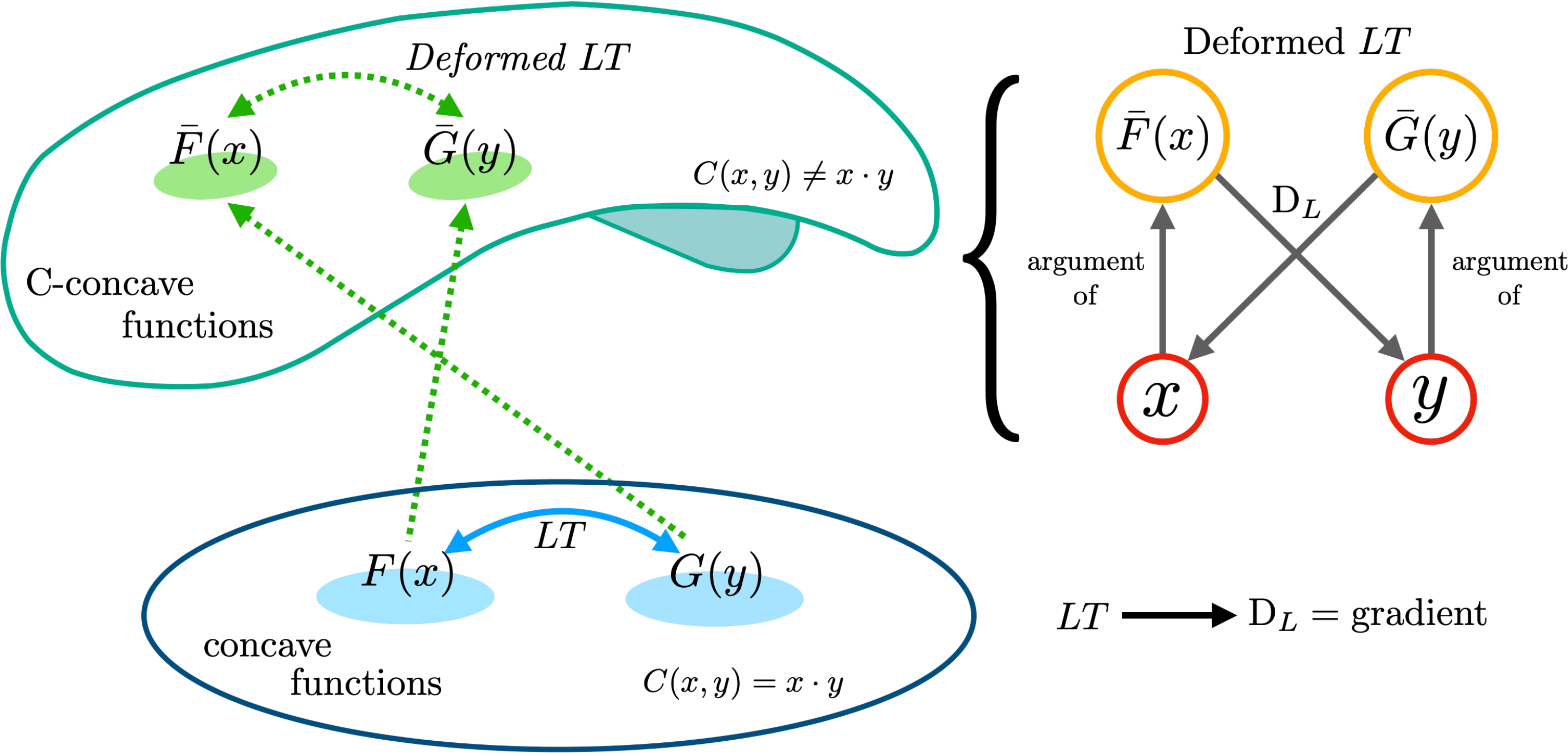}
    \caption{A graphical representation of the Legendre transform and its deformations. \textbf{Left:} While the standard Legendre transform acts on concave duals, the deformed one acts between $C$-concave functions. Each transform brings elements of one space to the other. 
    Please note that while Section~\ref{sec:prelim} presents the classical view of Legendre transforms acting over convex functions, the rest of this work follows Ref. \cite{wong2018logarithmic} in focusing on concave functions.
    \textbf{Right:} The symmetry that governs the algebraic relationships between convex dual functions and dual coordinates, which is mediated by the Legendre derivative operator $\text{D}_\text{L}$, which differs from the standard Euclidean gradient when the transform is deformed.}
    \label{fig:model2}
\end{figure*}

A more principled view comes from an algebraic perspective as follows. If $F(x)$ with $x=(x_1,\dots,x_n)\in\mathbb{R}^n$ is a strictly convex function (i.e., its Hessian is positive-definite), then the partial derivative $y_i(x):=\partial F/\partial x_i(x)$ is a monotonous function of $x_1,\dots,x_n$ for $i=1\dots n$. This means that there exists an isomorphism between $x$ and $y=(y_1,\dots,y_n)$; said differently, there exist mappings $y_i(x)$ and $x_k(y)$ that transform one into the other. Using these mappings, it would be natural to consider the possibility of reparametrizing $F$ in terms of $y$ instead of $x$. 
However, instead of focusing on such reparametrization, an elegant move is to consider instead the function $G(y) = x\cdot y - F\big(x(y),y\big)$. Interestingly, the resulting pair $F(x)$ and $G(y)$ exhibit the following symmetry:
\begin{equation}\label{eq:LLL}
    \frac{\partial G}{\partial y_k} = x_k,
    \quad
    \frac{\partial F}{\partial x_i} = y_i.
\end{equation}
Useful properties of this transformation are that it preserves convexity (i.e., the transform of a convex function results into another convex function) and it is an `involution', that is, the Legendre transform of the transform of a convex function is the function itself. 
The symmetry of these relationships is graphically represented in the right-hand side of Figure~\ref{fig:model2}. 

Overall, one can think of the Legendre transform as acting on two inputs, $x$ and $F$, and providing two outputs: the dual variable $y$ and the convex conjugate $G$~(similarly, the Fourier transform of a time series $F(t)$ can be thought of as giving two outputs too: the spectrum of amplitudes $G(s)$ (analogous to the conjugate function) and the frequency domain $s$ itself (analogous to the dual variable)).  
Pairs of convex functions $\{F,G\}$ satisfying Equation~\eqref{eq:LLL} are known 
as \textit{convex duals}, with $\{x,y\}$ being known as \textit{dual variables}. Additionally, convex functions and their duals satisfy the Fenchel inequality $F(x) + G(y) \geq x\cdot y$. 
The multiple useful properties of  Legendre duals are leveraged in various areas of mathematics and engineering, particularly in convex optimization~\cite{boyd2004convex}.

A more general definition of the Legendre transform of a convex function is given by
\begin{equation}\label{eq:legendre_basic}
    G(y) = \sup_{x}\{ C(x,y) - F(x) \}.
\end{equation}
This definition applies even when $F$ is not everywhere differentiable, and recovers the above procedure for the case where $C(x,y)=x\cdot y$. For other choices of $C$, this opens the door to so-called ``deformed'' Legendre transforms, which play an important role in optimal transport theory~\cite{villani2009optimal}. Interestingly, dual functions according to these generalized Legendre transforms satisfy relationships analogous to Equation~\eqref{eq:LLL}, but where the role of the Euclidean gradient is replaced by a `Legendre derivative' operator $\text{D}_\text{L}$, which is formally defined in Section~\ref{sec:generalized_legendre}. 
The goal of this paper is to explore the implications of such deformations of the Legendre transforms for physical systems. 

\section{Legendre Transform in Information Geometry}
\label{sec:legendre_in_geo}

In this section we present the key role of the Legendre transform in statistical manifolds. For this purpose, Section~\ref{sec:dual_geometry} first introduces the necessary background about information geometry to the unfamiliar reader. Then, Section~\ref{sec:dually_flat_geometry} explains how the standard Legendre transform describes the geometry of dually flat spaces, which are naturally associated with the Kullback--Leibler divergence and the Shannon entropy. Building on this, Section~\ref{sec:divergences_as_tools_geometries} then presents how other divergences lead to more general geometries, and Section~\ref{sec:generalized_legendre} develops how generalized Legendre transforms are a natural way to build and describe them. 
Please note that hereafter we use Einstein's summation convention for convenience of the notation.

\subsection{The Dual Structure of Statistical Manifolds}
\label{sec:dual_geometry}

Our exposition is focused on statistical manifolds $\mathcal{M}$ whose elements are probability distributions $p_\xi(s)$, with $s\in\boldsymbol S$ being the possible events accounted for by the probability distribution and $\xi\in \pazocal{O}\subset \mathbb{R}^d$ with $\pazocal{O}$ an open subset of a set of parameter values. The geometry of such statistical manifolds is determined by two structures: a metric tensor $g$ 
and a torsion-free affine connection pair $(\nabla, \nabla^{*})$ that are dual with respect to $g$. Intuitively, $g$ establishes norms and angles between tangent vectors and, in turn, establishes curve length and the \emph{shortest} curves. 
On the other hand, the affine connection establishes covariant derivatives of vector fields establishing the notion of parallel transportation between neighboring tangent spaces, which defines what is a \emph{straight} curve.

Traditional Riemannian geometry is built on the assumption that the shortest and the straightest curves  {locally} coincide, which is pivotal to the development of general relativity. This assumption leads to the study of metric-compatible Levi--Civita connections, as its geodesics are locally distance-minimizing and satisfy $\nabla=\nabla^*$ {and are, hence, completely determined} from the metric. However, modern approaches motivated in information geometry~\cite{amari2021information} and 
gravitational theories~\cite{Vitagliano:2010sr,Vitagliano:2013rna} consider more general scenarios, where connections may not be derivable from the metric. In such geometries, the parallel transport operator $\Pi:T_p\mathcal{M}\to T_q\mathcal{M}$ and its dual $\Pi^*$~(the dual transport operator acts on cotangent vectors and is defined by the condition of guaranteeing
$g_q (\Pi V,\Pi^{*} W)=g_p (V,W)$ for all $W\in T_p\mathcal{M}$ and $V\in T^*_p\mathcal{M}$) 
induced by $\nabla$ and $\nabla^*$, respectively might differ. 
The departure of $\nabla$ and $\nabla^*$ from self-duality can be shown to be proportional to Chentsov's tensor, which allows for a single degree of freedom traditionally denoted by $\alpha \in \mathbb{R}$~\cite{amari2021information}. Put simply, $\alpha$ captures the degree of asymmetry between short and straight curves, with $\alpha=0$ corresponding to metric-compatible connections where $\nabla=\nabla^*$.

An important property of the geometry of a statistical manifold ($\mathcal{M},g,\nabla, \nabla^{*}$) is its curvature, which can be of two types: the (Riemann--Christoffel) metric curvature or the curvature associated to the connection. Both quantities capture the distortion induced by parallel transport over closed curves, the former with respect to the Levi--Civita connection and the latter with respect to $\nabla$ and $\nabla^*$. In the sequel, we use the term \textit{curvature} to refer exclusively to the latter type. Statistical manifolds with zero curvature (equivalently, manifolds where it is possible to find a coordinate chart pair under which the connections and its dual vanish for any point of the manifold) are said to be dually flat.

\subsection{Dually Flat Geometry, Bregman Divergences, and the Legendre Transform}
\label{sec:dually_flat_geometry}

The geometry of Riemannian manifolds is typically formulated in terms of a single set of local coordinates. However, the fact that non-Riemannian manifolds have two dissimilar affine connections $\nabla$ and $\nabla^*$ makes it more natural to describe their geometry in terms of two dual coordinates $\xi$ and $\eta$~\cite{amari2021information}. Specifically, while in Riemannian geometry orthogonality can be assessed between the different dimensions of a single set of coordinates, in statistical manifolds it is more fruitful to consider orthogonality between elements of the primal $\xi$ and dual coordinates $\eta$~\cite{amari2001information,PhysRevResearch.3.033216}. A standard example of dual coordinates in a statistical manifold is where $\xi$ corresponds to the natural parameters of an exponential family distribution and $\eta$ corresponds to the corresponding expectation values. 
In the sequel, we follow Schouten's notation in which upper indices are reserved for dual coordinates, i.e.,
\begin{equation}\label{eq:dual_rep}
    \partial_{i}=\frac{\partial}{\partial \xi^i} \quad \text{and} \quad \partial^{i}=\frac{\partial}{\partial \eta_{i}}.
\end{equation}
Under this notation, $\partial_{i}$ gives rise to a basis for the tangent space $T_{p}\mathcal{M}$, while $\partial^{i}$ is related to a natural dual basis of the cotangent space $T^*_{p}\mathcal{M}$. 

A Riemannian metric is always ``locally flat'', i.e., it can be brought down to its signature (a Kronecker delta) at a given point $p \in \mathcal{M}$ by choosing an appropriate coordinate chart. It is not guaranteed, however, that such a chart would preserve the delta at a neighborhood of $p$; finding a chart that satisfies this property globally is the hallmark of a flat geometry. Analogously, affine geometries are also locally flat when considering its dual entry, therefore satisfying  $g(\partial_{i},\partial^{j})=\delta_{i}^{j}$ for an appropriate pair of primal and dual coordinate charts $\{\xi^{i}, \eta_{i}\}$ at some point $p$. In a similar fashion, this property in general only holds locally; dually flat geometries are characterized by the fact that one can find a pair of coordinates that satisfies this condition of orthogonality on the whole manifold~(under these coordinate charts, one can show that both the connections and its dual are vanishing, hence the term \textit{dual flatness}).  

For an orthogonal pair $\{\xi,\eta\}$ of a given dually flat manifold, the gradients of the mappings $\xi\mapsto\eta$ and $\eta\mapsto\xi$ are both symmetric. To confirm this, let us first note that 
\begin{equation*}
    g_{ij}=g( \partial_{i}\eta_{k}\partial^{k}, \partial_{j}) = \partial_{i}\eta_{k}g( \partial^{k}, \partial_{j})= \partial_{i}\eta_{k} \delta_{j}^{k}= \partial_{i}\eta_{j},
\end{equation*}
where the first equality follows from the chain rule of derivatives $\partial_{i}=\partial_{i}\eta_{k}\partial^{k}$. 
Then, using the fact that Riemannian metrics are always symmetric, one can see that $\partial_{i}\eta_{j}=g_{ij} = g_{ji} = \partial_{j}\eta_{i}$. 
A similar derivation shows that $g^{ij} = \partial^{i}\xi^{j}$, and hence $\partial^{i}\xi^{j}=\partial^{j}\xi^{i}$~(note that $g_{ij}=\partial_{i}\eta_{j}$ and $g^{ij}=\partial^{i}\xi^{j}$ is consistent with the fact that for orthogonal coordinates  $g(\partial^i,\partial_k)=g^{ij}g_{jk}=\delta_{k}^{i}$).

There is an intimate relationship between an orthogonal pair of coordinates in a dually flat manifold and the Legendre transform. 
To see this, we first note that the symmetry of the Jacobian of $\xi\to\eta$ implies the existence of a closed 1-form $\mathrm{d} \omega = 0$, and this---via Poincare Lemma---implies in turn the existence of a scalar potential $\psi\in C^{\infty}$ that satisfies 
\begin{equation}\label{eq:pro1}
    \eta_{i} = \partial_{i}\psi \quad \text{and} \quad  
    g_{ij} = \partial_{i}\partial_{j}\psi.
\end{equation}
Note that the second condition, combined with the fact that $g_{ij}$ is positive-semidefinite, implies that $\psi$ is convex. By a similar line of reasoning, the symmetry of $g^{i,j}$ induces a dual convex potential $\varphi$ that satisfies 
\begin{equation}\label{eq:pro2}
    \xi^{i}= \partial^{i}\varphi
    \quad\text{and}\quad
    g^{ij} = \partial^{i}\partial^{j}\varphi.
\end{equation}
Furthermore, a direct calculation shows that the dual potentials $\psi(\xi^1 , ..., \xi^n )$ and $\varphi(\eta_1 , ..., \eta_n)$ always satisfy $\dd\{ \psi + \varphi - \xi^{i}\eta_{i} \}= 0$. This implies that, modulo an unimportant constant, the following relationship holds over any dually flat manifold~(Equation \eqref{eq:pre-legendre} holds on any manifold but \emph{only locally}; in contrast, dually flat spaces are a special case in which dual potentials $\varphi,\psi$ that satisfy Equations~\eqref{eq:pro1} and \eqref{eq:pro2} can be defined over the whole manifold):
\begin{equation}\label{eq:pre-legendre}
    \psi + \varphi - \xi^{i}\eta_{i}= 0.
\end{equation}

Let us now consider the behavior of Equation~\eqref{eq:pre-legendre} on dually flat spaces when the coordinates and potentials are evaluated at different points of the manifold. For this, let us denote as $\xi(p)$ and $\eta(q)$ the coordinates and dual coordinates of $p,q\in\mathcal{M}$, respectively, and define the so-called \textit{Bregman divergence} $\pazocal{D}$ as
\begin{equation}\label{eq:Bregman}
    \pazocal{D}(p||q) :=  \varphi\big(\eta(p)\big) + \psi\big(\xi(q)\big) - \xi^{i}(q)\eta_i (p).
\end{equation}
Then, the differential of the mapping $q \mapsto \pazocal{D}(p_0||q)$ is
\begin{align}
    \label{eq:diff_divmap}
    \dd\big\{ \pazocal{D}(p_0,q)\big\}
    &= \Big( \partial_{i}\psi\big(\xi(q) \big) - \eta_{i}(p_0)  \Big)\mathrm{d}\xi^i (q) \nonumber \\
    &= \big(\eta_{i}(q)- \eta_{i}(p_0)\big)\mathrm{d}\xi^i (q).
\end{align}
From this, and considering that $\pazocal{D}$ by definition is a difference between a linear and two convex functions, one can verify that this mapping attains its unique minimum when $q= p_0$. Interestingly, at this minimal value one recovers Equation~\eqref{eq:pre-legendre}, which implies that $\pazocal{D}=0$. This shows that Bregman divergences are non-negative.

These results suggest an alternative definition for $\varphi$ and $\psi$, conceiving them as a maximum of the following maps:
\begin{align}
    \varphi(\eta(p)) &= \max_{q\in \mathcal{M}} \left \{  \xi^{i}(q)\eta_i (p) -\psi (\xi(q)) \right \},\\
    \psi(\xi(p))  &= \max_{q\in \mathcal{M}} \left \{  \eta_{i}(q)\xi^i (p) -\varphi (\eta(q)) \right \}.
\end{align}
This reveals that the orthogonal coordinate pair is always dual in the Legendre sense, or equivalently, that dual flatness implies that the potentials are convex duals. This property generalizes the well-known Legendre duality between the natural and expectation parameters of an exponential family~\cite{amari2000information}, showing that the same holds of any coordinate pair as long as they satisfy local flatness.

\subsection{Divergences as a General Tool to Establish Geometries}
\label{sec:divergences_as_tools_geometries}

This subsection explains how divergences, such as the one introduced in Equation~\eqref{eq:Bregman}, can be used as a convenient tool to establish a geometry on a statistical manifold~(\cite{amari2010information}, Section~4). Importantly, this approach does not lack generality, as any geometry can be expressed from an appropriate divergence~\cite{eguchi1983second,matumoto1993any,ay2015novel}. 

Divergences are a general class of functions that assess the dissimilarity of their arguments. More specifically, a divergence is a smooth,
distance-like function $\pazocal{D}[x;x']$ that satisfies $\pazocal{D}[x;x']\ge 0$ and vanishes only when $x=x'$. Divergences are more general---hence weaker---notions than distances, as they do not need to be symmetric in their arguments and may not respect the triangle inequality. Of the various types of divergences explored in the literature~\cite{Liese2006Divergences}, two are particularly important: \textit{$f$-divergences} (which are monotonic with respect to coarse-grainings of the domain of events $\boldsymbol S$~\cite{Amaridivergences}) and Bregman divergences (studied in the previous section).

Let us show how divergences can be used to establish metrics and connections over manifolds.
For this, let us use the shorthand notation $\pazocal{D}[\xi;\xi'] := \pazocal{D}(p||q)$ when expressing $\pazocal{D}$ in terms of coordinates $\xi=\xi(p)$ and $\xi'=\xi(q)$. Then, the Riemannian metric of the manifold is recovered from the second-order expansion of the divergence as follows:
\begin{equation}
\label{eq:FisherMetric}
  g_{ij}(\xi) 
  = \left \langle \partial_{i},
  \partial_{j} \right \rangle 
  = - \partial_{i , j'}
  \pazocal{D}[\xi;\xi'] \big|_{\xi=\xi'}~,
\end{equation}
which is positive-definite due to the non-negativity of $\pazocal{D}$. 
This construction leads to the \emph{Fisher's metric}, which is the unique metric that emerges from a broad class of divergences~(\cite{amari2010information}, Th.~5), with this being closely related with Chentsov’s theorem~\cite{chentsov1982statistical,ay2015information,van2017uniqueness,dowty2018chentsov}.
Similarly, connections emerge at the third-order expansion of the
divergence as follows:
\begin{subequations}
\label{eq:connections}
\begin{align}
\label{eq:C1}
  \Gamma_{ijk}(\xi) & = \left \langle \nabla_{\partial_{i}}
  \partial_{j} ,\partial_{k} \right \rangle 
  = -\;\left . \partial_{i ,j} \partial_{k'} \pazocal{D}[\xi;\xi'] \right|_{\xi=\xi'}\!,\\
\label{eq:C2}
  \Gamma_{ijk}^{*}(\xi) & = \left \langle \nabla_{\partial_{i}}^{*}
  \partial_{j} ,\partial_{k} \right \rangle 
  = -\left . \partial_{k} \partial_{i',j'} \pazocal{D}[\xi;\xi'] \right|_{\xi=\xi'}\!.
\end{align}
\end{subequations}
In summary, Fisher's metric is insensible the choice of divergence but the resulting connections are,
and therefore the effects of a particular $\pazocal{D}$ manifest only at the third order. 

Bregman divergences always give rise to flat geometries, as for them, $\partial_{i ,j} \partial_{k'} \pazocal{D}[\xi;\xi']=\partial_{k} \partial_{i',j'} \pazocal{D}[\xi;\xi']=0$, and therefore other types of divergences are needed in order to establish curved non-Riemannian geometries. As mentioned in Section~\ref{sec:dual_geometry}, the deviation of a given connection $\nabla$ from its corresponding metric-compatible (i.e., Levi--Civita) counterpart can be measured by $\alpha T$, where $T$ corresponds to the invariant \textit{Amari--Chensov} tensor~\cite{cencov2000statistical,amari1982differential} and $\alpha \in \mathbb{R}$ is a free parameter. The invariance of $T$ implies that the value of $\alpha$ entirely determines the connection, and the corresponding geometry can be obtained from a divergence of the form~\cite{PhysRevResearch.3.033216}
\begin{equation}\label{eq:alpha_div}
\pazocal{D}_{\alpha}(p || q)= \frac{4}{1-\alpha^2} \int_{\boldsymbol S}
  \left( 1 - p^{\frac{1-\alpha}{2}}(s)q^{\frac{1+\alpha}{2}}(s) \right) \dd\mu(s)~,
\end{equation}
which is known as \emph{$\alpha$-divergence}. As important particular cases, if $\alpha=0$ then $\pazocal{D}_{\alpha}$ becomes the square of Hellinger's distance, and if $\alpha=\pm1$ then it gives the well-known Kullback--Leibler divergence. 
Furthermore, it can be shown that the Kullback--Leibler divergence is a Bregman divergence, which in turn implies that for those cases the resulting geometry is flat. This illustrates the fact that being Riemannian (i.e., $\alpha=0$) and Euclidean ($\alpha=\pm1$) are independent features of a geometry.

We finish this subsection by noting that multiple divergences can give rise to the same geometry. 
A one-to-one relationship between divergence and geometries is obtained when considering \textit{conformal-projective} equivalent classes of divergences, which are related both via conformal and projective transformations. For a more detailed explanation, we refer the interested reader to Ref.~\cite{PhysRevResearch.3.033216}, Sec. 2-D.

\subsection{Generalized Legendre Transforms as a Natural Way to Describe Curved Manifolds}
\label{sec:generalized_legendre}

Sections \ref{sec:dually_flat_geometry} and \ref{sec:divergences_as_tools_geometries} clarified the intimate relationship that exists between dually flat manifolds, Bregman divergences, and the Legendre transform. Here we explain how these relationships are altered in more complex geometries.

In curved geometries it is impossible to construct dual potentials that satisfy Equation~\eqref{eq:pre-legendre} on the whole manifold. This impossibility is a symptom of the fact that the divergence that gives rise to this geometry, e.g., the $\alpha$-divergence given in Equation~\eqref{eq:alpha_div}, is not a Bregman divergence, but only an $f$-divergence~\cite{Amaridivergences}. 
To better understand the nature of the $\alpha$-divergence, let us consider in detail its relationship with Bregman divergences. Bregman divergences, as given in Equation~\eqref{eq:Bregman}, can also be expressed as 
\begin{align}
\pazocal{D}_\Phi[\xi;\xi']
= \Phi(\xi') - \Phi(\xi) - \textrm{D}\Phi(\xi)\cdot(\xi'-\xi).
\end{align}
Hence, $\pazocal{D}_\Phi[\xi;\xi']$ measures how convex the function $\Phi$ is at $\xi$ in the direction of $\xi'-\xi$~(this also explains the asymmetry that exists in the arguments of a Bregman divergence) and exploits the fact that a first-order approximation of a convex function always underestimates its value (i.e., that $\Phi(\xi') \geq \Phi(\xi) + \textrm{D}(\xi)\cdot(\xi'-\xi)$, where $\textrm{D}$ is the Euclidean gradient). Interestingly, such a first-order approximation can also be built on an intermediate point between $\xi$ and $\xi'$, which leads to
\begin{equation}\label{eq:ineq2}
\frac{1-\alpha}{2} \Phi(\xi) + \frac{1+\alpha}{2} \Phi(\xi') 
\geq 
\Phi\left( \xi_\alpha \right),
\end{equation}
where $\xi_\alpha=\frac{1-\alpha}{2} \xi + \frac{1+\alpha}{2} \xi'$, with $\alpha\in (-1,1)$ being a one-dimensional parameter that regulates how close $x_\alpha$ is to $\xi$ and $\xi'$. 
This inequality leads to a family of divergences~\cite{zhang2004divergence} indexed by $\alpha$, given by
\begin{equation}\label{eq:phi_div}
    \pazocal{D}_\Phi^{(\alpha)}[\xi;\xi']
:=\frac{4}{1-\alpha^2} \left[ 
\frac{1-\alpha}{2} \Phi(\xi) + \frac{1+\alpha}{2} \Phi(\xi') 
-\Phi\left( \xi_\alpha \right)
\right],
\end{equation}
where the factor $4/(1-\alpha^2)$ is introduced so that the limit $\lim_{\alpha\to1}\pazocal{D}_{\Phi}^{{}_{(\alpha)}} = \pazocal{D}_{\Phi}$ gives a Bregman divergence. In particular, if $\Phi(\xi) = \sum_i e^{\xi_i}$ then $\pazocal{D}_\Phi^{{}_{(\alpha)}}$ becomes the $\alpha$-divergence. 
Importantly, divergences of the form of Equation~\eqref{eq:phi_div} with $\alpha\neq \pm 1$ are not Bregman divergences (as they cannot be expressed in terms of convex conjugates as in Equation~\eqref{eq:Bregman}), and hence they do not lead to flat geometries (see Section~\ref{sec:divergences_as_tools_geometries}).

Fortunately, recent results suggest a way to express non-Bregman divergences in terms of generalized Legendre transforms~\cite{wong2018logarithmic}. The generalized Legendre transform is based on a \textit{link function} (Link functions are typically used as cost functions driving optimization problems in the literature focused on optimal transport~\cite{villani2009optimal}) corresponds to a smooth function $C: \mathcal{M}\times \mathcal{M} \to \mathbb{R}$, that connects generalized potentials $\varphi$ and $\psi$ via the following relationship:
\begin{equation}
\psi (\xi) + \varphi (\eta) - C(\xi,\eta) = 0,
\end{equation}
{which holds for all $(\xi,\eta)$ pairs belonging to the $C$-superdifferential of $\psi$. In this manner, $\eta$ can be interpreted as the $C$-supergradient of $\psi$ at $\xi$~\cite{pal2018exponentially}}. Put differently, for a given link function $C$, a pair of generalized potentials are functions $\varphi,\psi$, which are related via a generalized Fenchel--Lengendre $C$-transform as follows: 
\begin{subequations}
\begin{align}
\varphi\big(\xi(p)\big) &= \inf_{q\in \mathcal{M}} \left \{ \psi \big(\eta(q)\big) - C\big(\xi(p),\eta(q)\big)  \right \}, \label{eq:c-transform}\\
\psi\big(\eta(q)\big) &= \inf_{p\in \mathcal{M}} \left \{ \varphi \big(\xi(p)\big) -  C\big(\xi(p),\eta(q)\big) \right \}.\label{eq:c-transform2}
\end{align}
\end{subequations}
Note that these equations use a different sign than Equation~\eqref{eq:legendre_basic}, which leads to the consideration of concave instead of convex functions. Arguments for adopting this choice are discussed in Ref.~\cite{wong2018logarithmic}.

Following the rationale that led to Equation~\eqref{eq:Bregman}, for a given function $C$ and $C$-conjugate potentials $\varphi,\psi$, one can define a \textit{generalized Bregman divergence} (This divergence is known as a $C$-divergence, recently introduced in the context of optimal transport~\cite{pal2018exponentially}), where $C$ refers to the corresponding cost function. Here we use another term to stress its relationship with key geometric notions, given by
\begin{equation}\label{eq:gen_bregman}
    \mathcal{D}(p || q) = C\big(\xi(p),\eta(q)\big) - \varphi \big(\xi(p)\big) - \psi \big(\eta(q)\big) .
\end{equation}
Equations~\eqref{eq:c-transform} and \eqref{eq:c-transform2} 
imply that $\mathcal{D}(p || q)\geq 0$, with equality if and only if $p=q$. Interestingly, while the metric induced by generalized Bregman divergences is the Fisher metric,  Equations~\eqref{eq:C1} and \eqref{eq:C2} imply that the connections are given by
\begin{subequations}
\label{eq:C_connections}
\begin{align}
  \Gamma_{ijk}(\xi) & = 
  -\;\left . \partial_{i ,j} \partial_{k'} C\big(\xi;\eta(\xi')\big) \right|_{\xi=\xi'}\!,\\
  \Gamma_{ijk}^{*}(\xi) & = 
  -\left . \partial_{k} \partial_{i',j'} C\big(\xi;\eta(\xi')\big) \right|_{\xi=\xi'}\!.
\end{align}
\end{subequations}
If $C(\xi,\eta)=\xi\cdot\eta$ then $\Gamma_{ijk}(\xi)=\Gamma_{ijk}^{*}(\xi)=0$, and hence curved geometries in this construction only arise from non-trivial link functions, i.e., from deformations of the Legendre transform.

For the dual geometries that arise from the $\alpha$-divergence, one can identify the corresponding link function following a two-step procedure. First, one applies a monotonous transformation that turns the $\alpha$-divergence into the R\'{e}nyi divergence \cite{renyi1976} of order $\gamma$ (Note that we follow Ref.~\cite{valverde2019case} in adopting a shifted indexing, thereby referring to $\gamma = n-1$ as the order of R\'{e}nyi's entropy, with $n\ge 0$ corresponding to the order in the standard definition):
\begin{align} 
\label{eq:Renyialpha}
    \mathcal{D}_{\gamma}(p || q) =
    \frac{1}{\gamma} \log\int_{\boldsymbol S} p^{\gamma + 1}(s)q^{-\gamma}(s) d\mu(s),
\end{align}
related to the $\alpha$ parameter of divergence~\eqref{eq:alpha_div} as $\alpha=-1+2\gamma$ and leveraging the fact that both divergences generate the same geometry, being part of the same conformal-projective equivalent class~(\cite{PhysRevResearch.3.033216}, Sec. 2-D).
Note that when $\gamma\to0$, $C$ tends to $\xi\cdot\eta$, and the R\'{e}nyi divergence tends to the Kullback--{Leibler} divergence. 
As a second step, one uses the fact that the R\'{e}nyi divergence can be expressed in terms of generalized convex conjugates~(\cite{wong2018logarithmic}, Th.~13), and hence it can be recovered as a generalized Bregman divergence as Equation~\eqref{eq:gen_bregman}, where the link function is given by
\begin{equation}\label{eq:log_link}
C(\xi, \eta) = \frac{1}{\gamma}\log(1+ \gamma \xi^k \eta_k),
\end{equation}
and the corresponding generalized potential is
\begin{align}\label{eq:def_varphi}
  \varphi_{\gamma}(\xi) &= \log \int_{\boldsymbol S} (1+\gamma \xi \cdot h(s))^{-\frac{1}{\gamma}}d\mu(s).
\end{align}
Furthermore, it has been shown that this non-trivial logarithmic link function---or, equivalently, the R\'{e}nyi divergence---gives rise to dual geometries of constant curvature~\cite{wong2018logarithmic}. Therefore, this divergence constitutes a natural first step in the exploration of statistical manifolds of more complex geometry.

To conclude, let us introduce the notion of \textit{Legendre derivative} (This 
  corresponds to the $C$-gradient in optimal transport theory (see, e.g.,~\cite{wong2018logarithmic})). For given generalized potentials $\varphi$ and $\psi$, the corresponding \textit{Legendre derivative} is the operator $\textrm{D}_\text{L}$ that satisfies
\begin{equation}\label{eq:leg_derivative}
    \mathrm{D}_\text{L}\varphi(\xi) = \eta
    \quad \text{and}\quad
    \mathrm{D}_\text{L}\psi(\eta) = \xi. 
\end{equation}
The functional form for $\textrm{D}_\text{L}$ is determined by the corresponding link function. For example, for the case of $C(\xi,\eta)=\xi\cdot\eta$, Equations~\eqref{eq:pro1} and \eqref{eq:pro2} show that $\mathrm{D}_\text{L}$ is given by the Euclidean gradient. In contrast, for a logarithmic link function as in Equation~\eqref{eq:log_link}, one can find that the corresponding (non-Euclidean) Legendre derivative acting on a smooth function $\varphi$ is given by
\begin{equation}
\label{eq:dual_coordinates}
    \mathrm{D}_\text{L}^{(\gamma)}\varphi
    =
    \frac{1}{1-\gamma \xi\cdot\mathrm{D}\varphi}\mathrm{D}\varphi,
\end{equation}
with $\mathrm{D}$ denoting the Euclidean gradient.

\section{Symplectic and K\"{a}hler Structures in Information Geometry}
\label{sec:sym_and_kahler}

This section studies the realization of symplectic structures in statistical manifolds. This naturally leads towards considering the complexification of statistical manifolds, which enables a new avenue to develop insights about the Legendre transform. Complex manifolds are `bigger' bundles that possess a richer structure benefited by greater symmetry. These complex structures are quintessential to physics, being related to the quantization of the spin and coherent states~\cite{kochetov19952},  entanglement~\cite{brody2001geometric}, string theory~\cite{gawedzki1992non}, and  K\"{a}hler oscillators~\cite{PhysRevD.67.065013,PhysRevD.71.089901}.

The reasoning pursued here is that by recasting manifolds as complex structures with a higher degree of symmetry, one can obtain a more detailed understanding of their geometry and their relationship with the deformed Legendre transform. 
To develop this idea, we first establish a parallel between statistical manifolds and phase spaces. In doing this, it is important to note that while in statistical manifolds the dual coordinates $\xi$ and $\eta$ usually refer to the same point, in phase spaces they typically refer to canonical pairs (e.g., position and momentum) and hence correspond to different dimensions. This naturally leads to the consideration of product manifolds of two times the dimensionality of the original one.

\subsection{Establishing Dynamics on Phase Space}

In analytical mechanics, the Legendre transform enables the derivation of the Hamiltonian formalism from the Lagrangian, a smooth function of $n$ generalized coordinates $q$, velocity $\Dot{q}$, and time $t$. 
By doing this, one trades $n$ second-order equations of motion for $2n$ first-order differential equations of the form
\begin{equation}\label{eq:HamiltonEOMs}
    \frac{\partial H}{\partial p_j}=\Dot{q}^j\; , \frac{\partial H}{\partial q^k}=-\Dot{p}_k~.
\end{equation}
Notice that the transformation $(q,p)\mapsto (p,-q)$ preserves the form of the above equations. This symmetry is a reflection of a rich mathematical structure that provides the foundations of classical mechanics, which we introduce in the rest of this subsection.

We start by reviewing the standard method to establish dynamics over a manifold based on the Hamiltonian formulation of classical mechanics, as described, for instance, in Refs.~\cite{mcduff2017introduction,woodhouse1997geometric,bates1997lectures}. For this, let us consider a \textit{phase space} $\mathcal{M}$ that describes the possible configurations of a system of interest. More specifically, each point in $\mathcal{M}$ has the form $z=(q^1,...,q^n ,p_1,...,p_n)$, with $(q^1,...,q^n) \in \mathbb{R}^n$ corresponding to a configuration manifold $Q$, and $(p_1,...,p_n) \in \mathbb{R}^n$ corresponding to its generalized conjugate momenta. 
Dynamics over the phase space $\mathcal{M}$ are established by a Hamiltonian $H:\mathcal{M}\to \mathbb{R}$ via the following equations of motion:
\begin{equation}
    \dot{z}= X_{H},
\end{equation}
where the Hamiltonian vector field is given by
\begin{equation}\label{eq:pre_symplectic}
    X_{H}=J \mathrm{D}^{(0)} H(z)~, 
    \quad 
    \text{with}
    \quad
    J:= 
    \begin{pmatrix}
    0 & \mathds{1}\\
    -\mathds{1} & 0
    \end{pmatrix}
\end{equation}
and $\mathrm{D}$ denotes the standard gradient (see Equation~\eqref{eq:dual_coordinates}).
In this way, dynamics are established flowing the integral curves of $X_H$. At any point $z\in\mathcal{M}$ there is a trajectory governed by the dynamics induced by the Hamiltonian, which is unique due to the linearity of the equations involved.

Above, the role of Equation~\eqref{eq:pre_symplectic}---which turns the Hamiltonian into a vector field---can be re-framed in a more principled manner via symplectic geometry~\cite{arnol2013mathematical} as follows.
A symplectic form $\omega$ is a 2-form on $\mathcal{M}$ that is closed ($d\omega=0$) and non-degenerate ($\forall v \neq0 \;\exists u: \omega(v,u)\neq 0$). 
On a \emph{symplectic manifold} (i.e., a manifold equipped with a symplectic form), the flow of the Hamiltonian $H$ can be defined as the vector field $X_H$ that satisfies the following relationship:
\begin{equation}
\label{eq:HamiltonianVF_def}
    -\dd H = \iota_{X_{H}}\omega\,,
\end{equation}
where $\iota_{X}\omega = \omega(X, \cdot)$ is the 1-form that results from the interior contraction of $\omega$. 
Above, $\dd H$ is the differential of $H$ and the sign corresponds to a convention in the definition of the symplectic form. 
The fact that $\omega$ is non-degenerate guarantees that one can always find a unique $X_H$ that satisfies Equation~\eqref{eq:HamiltonianVF_def}. 
Additionally, the closure of the symplectic form locally implies---by the Poincare Lemma---the existence of a \textit{tautological} 1-form $\theta$ (also known as the canonical 1-form or symplectic potential), which satisfies the condition $\omega = \dd\theta$. This coordinate-invariant expression for $\omega$ emphasizes its topological nature.

Symplectic manifolds belong to equivalent classes established via symplectomorphism (i.e., diffeomorphism, which preserves the symplectic form), which are equivalent to canonical transformation in the context of analytical mechanics. 
The symplectic form allows us to determine a vector field from a smooth function up to diffeomorphisms that preserve the symplectic form, i.e., $\pazocal{L}_{X_{H}}\omega =0$. 
Furthermore, the geometry of the phase space gives an account of important properties of the underlying system. Indeed, while an unconstrained system may be described by a phase space of the form $\mathcal{M} = \mathbb{R}^{2n}$, more complicated systems are usually reflected by more convoluted geometries. As a simple example, a pendulum is described as a phase space of the form of a cylinder, which has a flat internal geometry but a non-trivial topology. The next subsections explore the implications of phase spaces with non-zero curvature.

\subsection{Symplectic Structure under the Deformed Legendre Transform}

Section~\ref{sec:divergences_as_tools_geometries} shows that, from an information-geometric perspective, divergences can be used to determine the metric and connections of a manifold. In this subsection, we show how divergences also generate a symplectic 2-form, from which much of the insights from Hamiltonian mechanics can be inherited. This, in turn, allows us to study probability distributions in phase space and discuss the flow induced by divergences. Our results will show that the symplectic 2-form induced by the divergence on the phase space and the induced Hamiltonian dynamics are different from the ones induced on the product manifold when the geometry is curved---or equivalently, when the Legendre transform has been deformed. 

To start, let us introduce some terminology. We will contrast structures on the cotangent bundle of statistical manifolds with structures in the product manifold $\mathcal{M} \times \mathcal{M}$ made of pairs of the form $(p,q)$. The product manifold is often parameterized using dual coordinates as $(\xi,\eta):=\big(\xi(p),\eta(q)\big)$~(as a consequence, in this section $\xi$ and $\eta$ refer to different points in the manifold, unless it is explicitly specified to be otherwise). In addition, let us use the projection operators over the left and right elements, $\pi_\text{l}(p,q)=p$ and $\pi_\text{r}(p,q)=q$, to define the sub-manifolds 
$\mathcal{M}_{q} :=\pi_\text{l}^{-1}(p,q) = \mathcal{M}\times \{q \} \simeq \mathcal{M}$ and $\mathcal{M}_{p} :=\pi_\text{r}^{-1}(p,q) = 
\{p\}\times \mathcal{M} \simeq \mathcal{M}$. The diagonal of the product manifold will be denoted as $\Delta \subset \mathcal{M} \times \mathcal{M}$, being made by pairs of the form $(p,p)$.

Divergences are smooth functions mapping $\mathcal{M} \times \mathcal{M}$ into $\mathbb{R}$, and we are interested in the geometrical structure that such mappings induce. 
To investigate this, let us consider the canonical symplectic form $\omega_p$ on $T^{*}\mathcal{M}_{p}$, which can be expressed in terms of a local chart $(U,\xi^{k},\nu_{k})$ as
\begin{equation}
    \omega_{p} :=\dd \xi^j \wedge \dd \nu_j\,,
\end{equation}
with $\nu_{k}$ being the conjugate coordinate to $\xi^k$. Note that, thanks to Darboux's theorem~\cite{mcduff2017introduction}, such canonical pairs are guaranteed to always exist locally. Let us then recast the map presented in Equation~\eqref{eq:diff_divmap} as the symplectomorphism $\mathrm{L}_{\pazocal{D}}:\mathcal{M}\times\mathcal{M}\to T^{*}\mathcal{M}_{p}$ given by
\begin{equation}
    \mathrm{L}_{\pazocal{D}} : (\xi,\eta) \mapsto (\xi,\nu) =(\xi, \partial_{i}\pazocal{D}(\xi,\eta)\mathrm{d}\xi^i)~.
\end{equation}
As shown in~\cite{zhang2013symplectic,leok2017connecting}, this map induces---via the pull-back $\mathrm{L}_{\pazocal{D}}^{*}\omega_{p} = \omega_{\pazocal{D}}$---the following symplectic form on $\mathcal{M}\times \mathcal{M}$:
\begin{subequations}
\begin{align}
    \mathrm{L}_{\pazocal{D}}^{*}\omega_{p} &= \mathrm{L}_{\pazocal{D}}^{*}[\dd \xi^{i} \wedge \dd \nu_{i}] \\
    &= \dd \xi^{i} \wedge \dd \{\partial_{i}\pazocal{D}(\xi,\eta)\} \\
    &= \dd \xi^i \wedge (\partial_{i,k}\pazocal{D}(\xi,\eta)\dd \xi + \partial_{i}^{\hphantom{i}k'}\pazocal{D}(\xi,\eta)\dd \eta_k) \label{eq:pullback_2}\\
    &= \partial^{k'}_{\hphantom{j'}i}\pazocal{D}(\xi,\eta) \dd \xi^i \wedge \dd \eta_{k}\,,\label{eq:cool_symplectic}
\end{align}
\end{subequations}
where 
 the vanishing of the first expression~\eqref{eq:pullback_2} is a result of the commutativity of the second derivatives of the divergence. Note that $\partial^{k'}_{\hphantom{j'}i}\pazocal{D}(\xi,\eta)$ reduces to the Fisher metric when evaluated on $\Delta$ (i.e., when $\xi$ and $\eta$ are evaluated at the same element $p$), but is different otherwise. Importantly, the same symplectic form on $\mathcal{M}\times\mathcal{M}$ is obtained by pulling back the canonical symplectic form $\omega_{q} := \dd \eta_{k} \wedge \dd \lambda^k$ on $T^{*}\mathcal{M}_{q}$ (where $(\eta, \lambda)$ form a canonical pair) in an analogous fashion, using here the symplectomorphism $\mathrm{R}_{\pazocal{D}}:\mathcal{M}\times\mathcal{M}\to T^{*}\mathcal{M}_{q}$ given by
\begin{equation}
    \mathrm{R}_{\pazocal{D}} : (\xi,\eta) \mapsto (\eta,\lambda) =(\eta, \partial^{k}\pazocal{D}(\xi,\eta)\mathrm{d}\eta_k)\,.
\end{equation}

Now that the symplectic form given by Equation~\eqref{eq:cool_symplectic} has been identified as the natural one on $\mathcal{M}\times\mathcal{M}$, our next step is to investigate how is it influenced by the manifold's curvature. For this, note first that if the divergence $\pazocal{D}$ is a generalized Bregman divergence, then its associated symplectic form depends solely on the link function. In effect, a direct calculation shows that for this case
\begin{equation}
    \omega_{\pazocal{D}} = \partial^{k}_{\hphantom{j'}i}C(\xi,\eta) \dd \xi^i \wedge \dd \eta_{k}\,.
\end{equation}
This clarifies how, although identical on the cotangent bundle $T^*\mathcal{M}$, the symplectic structure induced by different divergences may differ on $\mathcal{M}\times \mathcal{M}$.

\subsubsection{R\'{e}nyi's Symplectic 2-Form and Flow}
While the dually flat geometry established by Bregman divergences leads to a symplectic form given by $\omega_{\pazocal{D}}=\dd \xi^i \wedge \dd \eta_i$, for $\gamma$-curved geometry the R\'{e}nyi divergence induces the following symplectic form:
\begin{equation}
    \label{eq:symplectic_form_Div}
    \omega_\pazocal{D} = 
    \frac{1}{1+\gamma \xi^i \eta_i}\left(\delta^{\hphantom{l}k}_{l}-\frac{\gamma \xi^k \eta_l}{1+\gamma \xi^i \eta_i} \right)\dd \eta_k \wedge \dd \xi^{l}\,.
\end{equation}
The coefficients of this symplectic form coincide with the metric tensor in Ref.~\cite{wong2018logarithmic} (Proposition 4), this time on the product manifold $\mathcal{M}\times\mathcal{M}$.

The symplectic form exhibited in Equation~\eqref{eq:symplectic_form_Div} is closed, as can be confirmed by a direct calculation leading to $\dd \omega_{\pazocal{D}}=0$. This, in turn, implies the local existence of a corresponding tautological 1-form via Poincare Lemma, as explained in the previous section. Similar to the derivation that led to Equation~\eqref{eq:symplectic_form_Div}, we define the canonical 1-form $\theta_{p} = \nu_{i}\dd \xi^i$ on $T^{*}\mathcal{M}_p$ and evaluate its pull-back onto $\mathcal{M}\times \mathcal{M}$, yielding
\begin{equation}
    \theta = \frac{1}{2}\frac{\eta_{i}\, \dd \xi^{i} -\xi^{i}\, \dd \eta_{i}}{1+\gamma \xi^k \eta_k}~. \label{eq:canonical_1form}
\end{equation}
This expression, hence, characterizes the 1-form emerging from connections that describe the projective-flat geometry induced by R\'{e}nyi's divergence.

As a last step, let us leverage the symplectic form $\omega_\pazocal{D}$ to evaluate the action of the smooth function $\pazocal{D}_\gamma$ on the product manifold $\mathcal{M}\times\mathcal{M}$. This function is of particular interest as it generates integral curves of constant $\pazocal{D}$, and hence the induced flow is closed within the diagonal $\Delta\simeq\mathcal{M}$. 
For this purpose, let us denote as $X_{\gamma}= X_{\gamma}^{i}\partial_{\xi^i} + X_{\gamma j}\partial_{\eta_j}$ the vector field generated by the observable $\pazocal{D}_\gamma$ and the corresponding symplectic form.
We are interested in the vector fields that preserve the symplectic form $\omega_{\pazocal{D}}$, i.e., the vector field $X_\gamma$ that satisfies  $\pazocal{L}_{X_{\gamma}}\omega_{\pazocal{D}}=0$, where $\pazocal{L}_{X_{\gamma}}\omega_{\pazocal{D}}$ denotes the Lie derivative of $\omega_\pazocal{D}$ in the direction of $X_\gamma$. Then, using Cartan's magic formula one can find that
\begin{equation} \label{eq:Lie_deriv}
    \pazocal{L}_{X_{\gamma}}\omega_{\pazocal{D}}=\iota_{X_{\gamma}}\dd \omega_{\pazocal{D}} + \dd (\iota_{X_{\gamma}}\omega_{\pazocal{D}})=\dd (\iota_{X_{\gamma}}\omega_{\pazocal{D}}),
\end{equation}
where the last equality is a consequence of the fact that $\omega_{\pazocal{D}}$ is closed. Therefore, $\pazocal{L}_{X_{\gamma}}\omega_{\pazocal{D}}$ vanishes only if $X_{\gamma}$ is Hamiltonian~\eqref{eq:HamiltonianVF_def}, i.e., if $X_H$ satisfies
$\iota_{X_{\gamma}}\omega_{\pazocal{D}} + \dd \mathcal{D}_{\gamma}=0$. 
One can then determine the R\'{e}nyi vector field via explicit evaluation of the interior product as follows:
\begin{subequations}
\begin{align}
    -\mathrm{d}\mathcal{D}_{\gamma} &= (\iota_{X_{\gamma}}g_{k}^{\hphantom{k}l}\mathrm{d}\eta_{l})\wedge \mathrm{d}\xi^{k} - g_{k}^{\hphantom{k}l} \mathrm{d}\eta_{l} \wedge (\iota_{X_{\gamma}}\mathrm{d}\xi^{k}) \\
    &= g_{k}^{\hphantom{k}l}(X_{\gamma l} \mathrm{d}\xi^k - X_{\gamma}^{k} \mathrm{d}\eta_{l}),
\end{align}
\end{subequations}
which results in a Hamiltonian flow generated by R\'{e}nyi's divergences of the form
\begin{equation}
    X_{\gamma k} = -g_{k}^{\hphantom{k}a}\partial_{\xi^a} \mathcal{D}_{\gamma}\,, \quad X_{\gamma}^{\hphantom{\gamma}k} = g_{\hphantom{k}a}^{k}\partial_{\eta^a} \mathcal{D}_{\gamma} \,.
\end{equation}
Then, the corresponding R\'{e}nyi vector field can be found to be equal to
\begin{subequations} \label{eq:RenyiVF}
\begin{align}
    X_{\gamma} &= g_{\hphantom{k}a}^{k}\partial_{\eta^a} \mathcal{D}_{\gamma}\partial_{\xi^k} - g_{k}^{\hphantom{k}a}\partial_{\xi^a} \mathcal{D}_{\gamma}\partial_{\eta_k} \\
     &= [\eta(p)-\mathrm{D}^{(\gamma)}_{\mathrm{L}}\psi (q)]_{k} \partial_{\xi^k} \nonumber \\
    &\quad \, - [\xi(p)-\mathrm{D}^{(\gamma)}_{\mathrm{L}}\varphi (q)]^{k} \partial_{\eta_k}\,,
\end{align}
\end{subequations}
where $\mathrm{D}^{(\gamma)}_\mathrm{L}$ is the Legendre derivative operator introduced in Section~\ref{sec:generalized_legendre}.

As mentioned above, this R\'{e}nyi flow is closed within the diagonal $\Delta$. Moreover, the above result implies that the flows on the diagonal follow the geodesic with respect to the primal and dual connections, which naturally satisfy Equation~\eqref{eq:leg_derivative}.
In this way, we gain a new understanding of what deforming the exponential family implies. The squared brackets in Equation~\eqref{eq:RenyiVF} imply that the set of points flowing along the integral curves at $X_{\gamma}$ correspond to enforcing the dual coordinate pair as the Legendre derivative of the potential at the diagonal. Hence, the Bregman limit (i.e., $\gamma \to 0$) leads to the dual parameterization of exponential families from regular Legendre transformation, whereas finite $\gamma\neq 1$ leads to the deformed family of distributions obtained from R\'{e}nyi's divergence~(\cite{wong2018logarithmic}, Section 4), which would describe the sets of points flowing along the integral curves at $X_{\gamma}$ and external points diverging away from it.

\subsection{Complexification of Statistical Manifolds}

This section discusses some fundamental aspects of complex geometry, followed by the complexification of statistical manifolds. Then, the next section focuses on the complex structure induced by the R\'{e}nyi divergence. For a more extensive treatment of the properties of complex manifolds, we refer the reader to Refs.~\cite{candelas1988lectures,bouchard2007lectures,nakahara2018geometry}.

A complex manifold can be depicted as a topological space that locally looks like $\mathbb{C}^n$. 
One way to try building a complex manifold would be to  consider a $2n$-dimensional real manifold, and then arrange a set of coordinates $\{x^{k}_\text{c}\}$ into complex combinations such as $x^{2k-1}_\text{c}+i x^{2k}_\text{c}$. Unfortunately, such an arrangement is not only arbitrary, but also, more importantly, it is coordinate-dependent. In effect, additional structure on the manifold is required for it to be `complexifiable'.

One way to build a complex manifold is via a tensor field $J_{a}^{\hphantom{a}b}$ of real components satisfying $J^2 =-1$, which provides a linear endomorphism $J:T_{p}\mathcal{M}\to T_{p}\mathcal{M}$. Notably, the diagonalization of such a tensor cannot be accomplished in a vector space of real values; hence, the coefficients of vectors in $T_{p}\mathcal{M}$ must be allowed to be complex-valued (i.e., $T_{p}^{\mathbb{C}}\mathcal{M}=T_{p}\mathcal{M}\otimes \mathbb{C}$). By arranging $2n$-local coordinates into complex coordinates $x^{k}+iy^{k}$, e.g., via $x^k=x^{2k-1}_\text{c}, y^k=x^{2k}_\text{c}$, one can express $J$ in complex coordinates as
\begin{equation}
    J= i\dd z^{a}\otimes \frac{\partial}{\partial z^a} - i\dd z^{\bar{a}}\otimes \frac{\partial}{\partial z^{\bar{a}}}\,.
\end{equation}
Hereon, $a$ and $\bar{a}$ are indices within $\{1,\dots,n\}$, with the bar being used to distinguish between holomorphic and anti-holomorphic components. 
The manifold $\mathcal{M}$ together with the tensor $J$ are known as an ``almost complex structure''. 
With the aid of $J$, such complexified $T_{p}\mathcal{M}$ can now be decomposed into holomorphic and anti-holomorphic parts via projection operators given by $[P_{(\pm)}]_{ a}^{\hphantom{a}b}=\tfrac{1}{2}(\delta_{a}^{\hphantom{a}b} \pm J_{a}^{\hphantom{a}b})$. These projection operators can be used to decompose any $k$-form into $(p,q)$-forms with $p+q=k$.

As suggested above, every complex manifold is also a real manifold but the converse does not always hold. A necessary and sufficient condition on $J$ to allow a real manifold to be a complex one is given by $N_{ab}\vphantom{N}^{c}=0$, where $N_{ab}\vphantom{N}^{c}$ stands for the Nijenhuis tensor given by~(note that the connections appearing from the covariant derivatives cancel out, which is why it is often found written in terms of partial derivatives in spite of being a tensor)
\begin{equation}
\label{eq:Nijenhuis}
    N_{ab}\vphantom{N}^{c}:= 2\Big(J_{a}\vphantom{J}^{d}\nabla_{[d} J_{b]}\vphantom{J}^{c}- J_{c}\vphantom{J}^{d}\nabla_{[d} J_{a]}\vphantom{J}^{c}\Big)\,,
\end{equation}
with squared brackets denoting the antisymmetrization of indices.

Up to this point, the complex manifold $(\mathcal{M},J)$ has not been equipped with a metric; in fact, a $J$-compatible metric may not exist (e.g., in Hopf manifolds). When such a metric does exist, this imposes the following compatibility conditions: 
\begin{equation}
g_{\mu \nu} J_{\rho}^{\hphantom{\rho}\mu} J_{\sigma}^{\hphantom{\sigma}\nu} = g_{\rho \sigma}
\quad
\text{and}
\quad
\nabla_{\mu}J_{\sigma}^{\hphantom{\sigma}\nu} = 0. 
\end{equation} 
The first condition above implies that the pure holomorphic and anti-holomorphic components of the metric vanish; hence, $ds^2 = g_{\mu \nu}dx^{\mu} \otimes dx^{\nu} = g_{a \bar{b}}dz^{a}\otimes d\bar{z}^{\bar{a}}$ is hermitian. The second condition enforces the vanishing of Nijenhuis tensor~\eqref{eq:Nijenhuis}, not only guaranteeing complexification, but also implying that the \textit{K\"{a}hler 2-form} given by
\begin{equation}\label{eq:zzz}
    k=\frac{1}{2}g_{\mu \nu} J_{\rho}^{\hphantom{\rho}\mu}\dd x^{\rho} \wedge \dd x^{\nu} = ig_{a \bar{b}}\dd z^a \wedge \dd\bar{z}^{\bar{b}}
\end{equation}
is closed, which serves as the manifold's symplectic form. In components, Equation~\eqref{eq:zzz} means that $\partial_{a}g_{b \bar{c}}=\partial_{b}g_{a \bar{c}}$ and $\partial_{\bar{b}}g_{a \bar{c}}=\partial_{\bar{c}}g_{a \bar{b}}$. Analogously as in~\eqref{eq:pro2}, these expressions can be locally integrated revealing the metric
\begin{equation} \label{eq:hermitian_metric}
g_{a \bar{b}}=\partial_a \partial_{\bar{b}}\pazocal{K}(z,\bar{z}),
\end{equation}
with $\pazocal{K}$ being a real-valued smooth function known as the \textit{K\"{a}hler potential}. This potential is not unique, as it is only determined up to the addition of a holomorphic and an anti-holomorphic function:
\begin{equation}
    \pazocal{K}(z,\bar{z}) \to \pazocal{K}(z,\bar{z}) + U(z) + \bar{U}(\bar{z})\,.
\end{equation}
Furthermore, $\pazocal{K}$ may not be globally defined~(if it were, the $\omega$ form would be exact and so its manifold's volume form implies the vanishing of its integral, violating the non-degeneracy condition for the metric). 
In this way, a Riemannian metric as well as the symplectic form are determined by $\pazocal{K}$, as $\omega = k = \tfrac{i}{2}\partial\bar{\partial}\pazocal{K}$
with $\{\partial,\bar{\partial}\}$ denoting the Dolbeault operators $\partial = \dd z\wedge \partial_{a}$ and $\bar{\partial} = \dd \bar{z}\wedge \partial_{\bar{a}}$.
The similarities between these expressions 
and the ones in Section~\ref{sec:generalized_legendre} and K\"{a}hler's are no coincidence, as $\pazocal{K}$ itself must convex. These similarities have been, in fact, the catalyst for the investigation of more intimate relations between the space of K\"{a}hler metrics and convexity~\cite{Bo_convexity} and various applications in the context of optimal transport~\cite{khan2020kahler}.

In statistical manifolds the fundamental object is its divergence $\pazocal{D}$, and therefore the constraints on the metric are ultimately enforced on $\pazocal{D}$. Hence, the conditions for complexification of a manifold translate into two conditions over the corresponding divergence~\cite{zhang2014divergence}:
\begin{enumerate}
    \item $\partial_{i,j'}\pazocal{D}=\partial_{j',i}\pazocal{D}$ on $\mathcal{M} \times \mathcal{M}$; \label{eq:condition1}
    \item $\partial_{i,j}\pazocal{D} + \partial_{i',j'}\pazocal{D} = \kappa\, \partial_{i,j'}\pazocal{D}$ for some $\kappa \in \mathbb{R}$. \label{eq:condition2}
\end{enumerate}
Above, the primed indices denote differentiation with respect to $y \in \mathcal{M}_{q}$ (as opposed to regular derivatives with respect to $x \in \mathcal{M}_{p}$). 
Although the first condition above is trivially satisfied when evaluated at the diagonal (as shown in Equation~\eqref{eq:FisherMetric}), it is not automatic for it to hold on the whole $\mathcal{M} \times \mathcal{M}$ manifold. Both conditions arise from the construction of an invariant arc element $\dd s^2$ from the symmetric and antisymmetric parts, given by
\begin{align}
\label{eq:invariantelem}
    \dd s^2 &= g_{\pazocal{D}}-i\omega_{\pazocal{D}} \\
    &= \partial_{i,j'}\pazocal{D}[x;y](\dd x^i \otimes \dd x^j + \dd y^i \otimes \dd y^j) \nonumber \\
    &\hphantom{=}\,+ i\partial_{i,j'}\pazocal{D}[x;y](\dd x^i \otimes \dd y^j - \dd y^i \otimes \dd x^j),
\end{align}
where $g_{\pazocal{D}}$ and $\omega_{\pazocal{D}}$ denote the metric and symplectic form induced by the divergence $\pazocal{D}$ on $\mathcal{M}\times\mathcal{M}$. Note that $\omega_{\pazocal{D}}$ is equivalent to the one derived in~\eqref{eq:canonical_1form}, while 
the components of $g_{\pazocal{D}}$ are expressed in Equation~\eqref{eq:hermitian_metric}. 
The second condition for the complexification of a statistical manifold is motivated by the fact that, if one is interested in expressing $\dd s^2$ as $\partial \bar{\partial}\pazocal{D}$, then the condition~\eqref{eq:condition2}
should be satisfied on $\mathcal{M}\times \mathcal{M}$ for $\kappa \in \mathbb{R}$. 

Importantly, divergences that can be expressed as in Equation~\eqref{eq:phi_div} 
for a given convex function $\Phi$ satisfy the conditions discussed above, and hence the geometries they induce are compatible with a complex structure~\cite{zhang2014divergence,khan2020kahler}. 
These divergences induce a geometry of constant scalar curvature given by $\kappa = \alpha_{-}\alpha_{+}$ with $\alpha_{+} = -\gamma$ and $\alpha_{-} = 1+\gamma$. Furthermore, $\Phi (\alpha_{+}x + \alpha_{-}y)$ serves as the local K\"{a}hler potential of the manifold. It is worth noting that $\gamma \to 0$ results in a vanishing $\pazocal{K}$ and thus cannot be defined. Indeed, $\gamma = 0$ is an excluded value for these expressions, and its limit should be previously worked out prior to complexification, as discussed in Ref.~\cite{zhang2014divergence}.

\subsection{Complex R\'{e}nyi Geometry under the Deformed Legendre Transform}

Let us now exploit the general results presented in the previous section to deepen our understanding of the geometry induced by the R\'{e}nyi divergence on statistical manifolds. The R\'{e}nyi divergence $\mathcal{D}_{\gamma}$ belongs to the family of divergences that can be expressed as in Equation~\eqref{eq:phi_div} using $\Phi(x)$ as given by
\begin{equation}
    \Phi(x) = \log \sum_{s\in \boldsymbol S} e^{x(s)}, \quad \text{with} \quad x(s)=:\log p(s).
\end{equation}
This means that the geometry that arises from the R\'{e}nyi divergence is susceptible to being complexified.  Furthermore, when evaluated on arguments that correspond to probability distributions (i.e., $x^a=\log p^{a}$ and $y^{a}=\log q^{a}$) then the first two terms in Equation~\eqref{eq:phi_div} vanish, and therefore the R\'{e}nyi divergence itself serves as the K\"{a}hler potential.

Let us now show that the two conditions for complexification discussed in the previous subsection are satisfied by product manifolds $\mathcal{M}\times \mathcal{M}$ endowed by a geometry induced by R\'{e}nyi's divergence. 
For this, we adopt complex coordinates $w^a = x^a + i y^a \in \mathbb{C}$ with $x^a = \log p^a$ and $y^a = \log q^a$ for $p,q\in\mathcal{M}$. 
Using these coordinates, one finds that
\begin{subequations}
\begin{align}
    -\frac{1}{\kappa}\mathcal{D}_{\gamma}(x,y) &= \log
    \sum_{a=1}^{n}\exp(\bar{\Gamma} w^a + \Gamma \bar{w}^a) \\
    &= \log z_{a} \bar{z}^{a},
\end{align}
\end{subequations}
where we are using the shorthand notations $\Gamma = \tfrac{1}{2}(\alpha_{-} + i \alpha_{+})$ and
$z^a = \exp (w\bar{\Gamma}^a)$.
In this manner, $\Phi (\alpha_{+}x + \alpha_{-}y)$ (or, equivalently, $\mathcal{D}_{\gamma}(x,y)$) can be identified as the K\"{a}hler potential for the product manifold.

The resemblance between the induced symplectic form in Equation~\eqref{eq:symplectic_form_Div} and the connections~\eqref{eq:canonical_1form} at the previous section to the well-known Fubini--Study metric and its connection are suggestive of the \textit{complex-projective spaces} $\mathbb{C}\mathrm{P}^n$~(for an overview on $\mathbb{C}\mathrm{P}^n$ spaces, please refer to Refs.~\cite{candelas1988lectures, bouchard2007lectures,nakahara2018geometry}). Unfortunately, complexification of the local charts does not preserve the functional form of the symplectic form given by Equation~\eqref{eq:symplectic_form_Div}, nor the canonical 1-form given by Equation~\eqref{eq:canonical_1form}. Nevertheless, special circumstances---such as $\gamma = 1$ and a restriction to the diagonal $\Delta$---do lead to $\mathbb{C}\mathrm{P}^n$ upon complexification.
Disregarding the pure holomorphic and anti-holomorphic functions of the divergence, the link function of the deformed Legendre transform can be directly read as the K\"{a}hler potential as follows:
\begin{equation} \label{eq:Kahler_Link}
    \pazocal{K}(z,\bar{z})=C(z,\bar{z}) = \log(1+ z_{a}\bar{z}^a)\,,
\end{equation}
hence generating the Fubini--Study metric given by
\begin{equation}
    g_{a\bar{b}} = \frac{1}{1+z_{a}\bar{z}^a}\left(\delta_{a\bar{b}}-\frac{z^a \bar{z}^b}{(1+z_{a}\bar{z}^a)} \right)\,.
\end{equation}
The case of complex dimension $n=\dim_{\mathbb{C}}\mathcal{M}=1$ (two real dimensions), that is, $\mathcal{M}= \mathbb{C}\mathrm{P}^1 \subset\mathbb{C}^2$, is of particular interest to physical systems. Indeed, from a group-theoretic perspective, this manifold corresponding to the coset group $SU(2)/U(1)$ (isomorphic to the Riemann sphere $S^2 \simeq \mathbb{C}\mathrm{P}^1$) is crucial for the formulation of spin coherent states~\cite{RevModPhys.62.867,kochetov19952} and the geometric quantization of the spin~\cite{woodhouse1997geometric}. In addition, $\mathbb{C}\mathrm{P}^1$ describes pure quantum states whose direct product enables a nice geometric formulation of many phenomena of interest, including entangled systems~\cite{brody2001geometric}.

The connection on this manifold corresponds to the canonical 1-form, which is now determined by its K\"{a}hler potential
\begin{equation}
    \pazocal{A}=\frac{i}{2}(\partial - \bar{\partial}) \pazocal{K}=
    \frac{i}{2}\frac{z_{a} \dd \bar{z}^a - \bar{z}_{a}\dd z^a}{1+z_{a}\bar{z}^a}
\end{equation}
via the Dolbeault operators~(here the index takes only one entry $a=1$, with trivial generalization to $\mathbb{C}\mathrm{P}^n$). 
Note that this gauge-field is consistent with the expression obtained for the connection 1-form found in Equation~\eqref{eq:symplectic_form_Div}.

Let us now show how a quantization of the 2-sphere restricts the allowed values for the R\'{e}nyi parameter $\gamma$. As Poincare's Lemma tells us, every closed form is \textit{locally} exact, and hence the existence of closed forms failing to be exact reflects some non-trivial aspect of the topology of the manifold. This feature is captured by cohomology classes $\pazocal{H}^{k}(\mathcal{M},\mathbb{R})$, whose members are closed yet \textit{globally} not exact $k$-forms. In this sense, the K\"{a}hler form belongs to $\pazocal{H}^{2}(\mathcal{M},\mathbb{R})$.
The single-valuedness of points on the manifold would require the $\omega_{\pazocal{D}}$ to belong to a cohomology class $\pazocal{H}^{2}(\mathcal{M},\mathbb{R})$. Therefore, its symplectic two-form must be an integer multiple of $\omega_{\pazocal{D}}$. Hence, the covariant derivative is $\nabla_a = \partial_a -i k \pazocal{A}_{z}$ with $k\in \mathbb{Z}$~(not to be confused with the manifold's complex dimension $n$), and the same holds for its anti-holomorphic counterpart. The holomorphic polarization~(see Appendix \ref{app_CP}) imposes the condition 
$\nabla_{\bar{a}} \psi = 0$ for $\psi$ wave function, a function whose squared module gives the probability density, closely resembling wave functions in quantum mechanics. This
results in
\begin{equation}
    \left(\partial_{\bar{z}} + \frac{k}{2}\frac{z^a}{1+z_{a}\bar{z}^a} \right)\psi = 0~.
\end{equation}
This implicit equation is solved by physical solutions $\psi_{\mathrm{phys}}$ of the form
\begin{equation}
    \psi_{\mathrm{phys}} = \exp\left( -\frac{k}{2}\log(1+z_{a}\bar{z}^a) \right)f(z)~,
\end{equation}
with $f(z)$ being a holomorphic function. The resulting probability density $|\psi_{\mathrm{phys}}|^2$ is given by
\begin{equation}
    \pazocal{P}(z)=\frac{|f(z)|^2}{(1+z_{a}\bar{z}^a)^{k}}~.
\end{equation}
The holomorphic function $f(z)$ can be expanded on the basis $\{1,z,z^{2},...,z^k\}$, as higher powers would imply $\pazocal{P}(z)$ to diverge; hence, a Hilbert space of finite dimension as $\psi_{\mathrm{phys}}$ is defined over the 2-sphere.

Just as holomorphic polarization for $\gamma=0$ results in exponential family distributions (Appendix~\ref{app_CP}), one recovers the R\'{e}nyi maximum entropy distributions as a polarization of the manifold for other values of $\gamma$.
Moreover, by identifying $\gamma = \tfrac{1}{k}$, one realizes (keeping the sign of $\gamma$) that $k \in \mathbb{Z}_{+}$ introduces the restriction  $\gamma \in (0,1]$, which corresponds to $\alpha \in (-1,1]$ and reflects a positive curvature, as discussed in Ref.~\cite{PhysRevResearch.3.033216}. Although ruled out by the polarization, it is interesting to note that considering $\gamma \not \in (0,1]$ would result in the manifold having hyperbolic topology and becoming non-compact, hence not being susceptible to complexification. These results establish $\gamma \in (0,1]$ as values of special physical significance: $\gamma =1$ for spin coherent states~\cite{kochetov19952}, worldline formalism~\cite{PhysRevD.103.036004}, K\"{a}hler oscillators~\cite{PhysRevD.67.065013,PhysRevD.71.089901}, and entanglement~\cite{brody2001geometric}, and other values in $\gamma \in (0,1]$ for systems described through the geometric quantization framework. Notably, this range does not include $\gamma =0$, which corresponds to conventional dually flat geometry and the \mbox{Shannon entropy.}

\section{Conclusions}
\label{sec:conclusion}

The Legendre transform, a fundamental piece of classic and contemporary physics, has a direct but non-trivial correspondence with the dually flat geometry of statistical manifolds induced by Shannon's entropy and the Kullback--Leibler divergence. This paper explores how deformations of the Legendre transform induce a departure from this regime and has multiple consequences on symplectic geometry and complexification. Taken together, these results provide some first steps towards a novel, rigorous, and encompassing understanding of physical systems that are not well-described by classic information-theoretic quantities.
The role of the Legendre transform on analytical mechanics differs from that in information geometry; in the latter, dual coordinates refer to different descriptions of the same point, whereas in the former, they refer to an isomorphism between the tangent and cotangent bundles.
In flat geometry the symplectic form of the cotangent bundle is equivalent to a canonical area form at the product manifold. In contrast, our results show that this equivalence is broken if the manifold is curved. 
Interestingly, this implies that a deformation of the regular Legendre transform results in the failure of the natural coordinates to form a canonical pair.
Furthermore, an analysis of the deformed symplectic form and flow that arises in curved manifolds reveals a new understanding of the family of maximum R\'{e}nyi entropy distributions, which are found to form
sets of points flowing along the integral curves of the flow. 
The departure of the symplectic form of the product manifold from the cotangent bundle provides a promising lead to study coupled physical systems,
with non-canonical coordinates---like the pair induced by the R\'{e}nyi geometry---being subjects of special interest. For instance, there have been studies on the consequences of deformations in the symplectic form in field theory~\cite{PhysRevE.79.011105}
and in $\mathbb{C}\mathrm{P}^n$ K\"{a}hler oscillators, where deformations to the symplectic structure via magnetic field are explored~\cite{PhysRevD.67.065013,PhysRevD.71.089901}. Other related phenomena have been studied in Fermi liquids under an external magnetic field, where the the magnetic field couples to Berry's curvature, deforming the symplectic form. Such deformations have been shown to have strong consequences for observables, as the invariant phase volume is modified via a topological invariant~\cite{Duval:2005vn,PhysRevLett.109.181602}.
An interesting avenue for future research is to investigate if there are divergences that can recapitulate these deformations, providing a mathematical scaffolding for the study of such 
systems.
In this work we have established a broad range of nonzero $\gamma$ values relevant from more than just a mathematical perspective.
Both symplectic topology and K\"{a}hler manifolds are sensitive to the topology rather than local changes in geometry. Furthermore, they are sensitive to the physical systems to which they now connect. 
In particular, our results show that $\gamma =1$ corresponds to a special case that is associated with the $\mathbb{C}\mathrm{P}^1$ manifolds relevant across various fields such as coherent states~\cite{kochetov19952}, worldline formalism~\cite{PhysRevD.103.036004}, K\"{a}hler \mbox{oscillators~\cite{PhysRevD.67.065013,PhysRevD.71.089901}} and entanglement~\cite{brody2001geometric}, to name a few. Via geometric quantization methods, our results show that holomorphic polarization leads to $\gamma \in (0,1]$. This reveals a further array of values of interest outside of the conventional $\gamma =0$ that characterizes the conventional dually flat Shannon systems.

The results presented here establish a first step in uncovering the consequences that the relationship between generalized Legendre transforms and curved statistical manifolds have for physical systems. We hope that this investigation may foster future work on these important implications, which may reveal other hidden threads connecting seemingly dissimilar approaches, such as the one revealed here relating non-Shannon entropies and non-canonical coordinates. Such investigations may lead towards a principled and unified understanding of physical systems that are not well-described by traditional approaches, providing solid foundations to support and guide some of today's effective but ad hoc procedures of analysis.

\acknowledgments{P.A.M. acknowledges support by JSPS KAKENHI Grant Number 23K168550001. J.K. acknowledges support by the Austrian Science Fund (FWF) project No. P 34994. F.R. was supported by the Fellowship Programme of the Institute of Cultural and Creative Industries of the University of
Kent.}

\appendix

\section{Complex Polarizations}
\label{app_CP}

This appendix illustrates the method of holomorphic polarization, which establishes an intimate relation between link functions and natural families. 
For a given K\"{a}hler manifold one can choose a polarization. A holomorphic polarization has the consequence that physical states are represented as holomorphic functions, thereby generalizing Bargmann--Segal's (Fock) spaces that are relevant to coherent states. The complex polarization is a condition determined by
\begin{equation}\label{eq:polarization}
    \nabla_{\bar{a}}\psi=\left(\partial_{\bar{a}}+\tfrac{1}{2}\partial_{\bar{a}}\pazocal{K}(z,\bar{z})\right)\psi =0~,
\end{equation}
where the connection is determined by the K\"{a}hler potential over the manifold. This polarization implies that the commutator $[\nabla_{\bar{a}},\nabla_{\bar{b}}]=0$, and hence the system described at~\eqref{eq:polarization} is integrable. Its general solution is given by
\begin{equation} \label{eq:pol_sol}
    \psi_{\mathrm{phys}}=\exp[-\tfrac{1}{2}\pazocal{K}(z,\bar{z})]\phi(z)~.
\end{equation}
In the context of statistical manifolds, $\pazocal{K}(z,\bar{z})$ corresponds to a link function $C(z,\bar{z})$. 
Therefore, Equation~\eqref{eq:pol_sol} corresponds to a natural family of distributions, e.g., the flat geometry $\mathbb{C}^{n}$ is described by $C(z,\bar{z})=z^{a}\bar{z}_{\bar{a}}$, which leads to the exponential family, whereas a link function of the form of Equation~\eqref{eq:Kahler_Link} yields R\'{e}nyi's natural family. The resulting physical Hilbert space is 
\begin{equation}
    \pazocal{H}_{\mathrm{phys}}=\left\{\phi(z) \left|  \int_{\mathcal{M}} \right. |\phi|^2 e^{-C(z,\bar{z})} \omega^n <\infty \right\}~,
\end{equation}
where $\omega^n$ denotes the manifold's volume form. In other words, one considers square-integrable global sections that are covariantly constant along $\nabla_{\bar{a}}$.

\bibliography{references}
\end{document}